\documentclass[conference,onecolumn]{IEEEtran}
\usepackage{cite}

\usepackage{subfigure}  % use for side-by-side figures
\usepackage[pdftex]{graphicx}   % need for figures

\usepackage[cmex10]{amsmath}
\newcommand{\eqs}{\begin{eqnarray*}}
\newcommand{\eqf}{\end{eqnarray*}}
\newcommand{\lef}{\left(}
\newcommand{\rig}{\right)}
\newcommand{\mas}{\begin{array}}
\newcommand{\maf}{\end{array}}
\newcommand{\deriv}{\textnormal{d}}

\begin{document}

\title{Modeling and tackling resistivity scaling in metal nanowires}

\author{\IEEEauthorblockN{Kristof Moors\IEEEauthorrefmark{1}\IEEEauthorrefmark{2},
Bart Sor\'ee\IEEEauthorrefmark{1}\IEEEauthorrefmark{3}\IEEEauthorrefmark{4} and
Wim Magnus\IEEEauthorrefmark{1}\IEEEauthorrefmark{3}}
\IEEEauthorblockA{\IEEEauthorrefmark{1}imec, Kapeldreef 75, B-3001 Leuven, Belgium}
\IEEEauthorblockA{\IEEEauthorrefmark{2}Institute for Theoretical Physics, KU Leuven, Celestijnenlaan 200D, B-3001 Leuven, Belgium}
\IEEEauthorblockA{\IEEEauthorrefmark{3}Physics Department, University of Antwerp, Groenenborgerlaan 171, B-2020 Antwerpen, Belgium}
\IEEEauthorblockA{\IEEEauthorrefmark{4}Electrical Engineering (ESAT) Department, KU Leuven, Kasteelpark Arenberg 10, B-3001 Leuven, Belgium\\Email: kristof@itf.fys.kuleuven.be}}

\maketitle

\begin{abstract}
A self-consistent analytical solution of the multi-subband Boltzmann transport equation with collision term describing grain boundary and surface roughness scattering is presented to study the resistivity scaling in metal nanowires. The different scattering mechanisms and the influence of their statistical parameters are analyzed. Instead of a simple power law relating the height or width of a nanowire to its resistivity, the picture appears to be more complicated due to quantum-mechanical scattering and quantization effects, especially for surface roughness scattering.
\end{abstract}

\section{Introduction}
An increasingly important factor for increasing the transistor density is the resistivity of the interconnects wiring all of them together. Current experimental data seems to indicate that the resistivity in metal thin films and wires scales inversely proportional to the width or diameter and that it is mainly driven by increased electron scattering due to grain boundaries (GBs) and surface roughness (SR) of the wire boundaries \cite{steinhogl2002size,josell2009size,chawla2011electron}. The wire dimensions have to scale down together with the transistor size, but because of the resistivity scaling there is an increase of many detrimental effects, such as heating, power consumption and signal delay. If the resisitivity increase cannot be kept under control, the interconnects become a major bottleneck hampering further downscaling.

While experimental data on the resistivity of metal wires with nanoscaled crosssection seems to be still in agreement with the commonly used Fuchs-Sondheimer and Mayadas-Shatzkes (MS) models, these models make use of fitting parameters and semi-classical approximations and do not provide further insight in the quantum-mechanical effects of scattering, confinement and the detailed structure of the GBs and SR \cite{fuchs1938conductivity,sondheimer1952mean,mayadas1970electrical}. In this work we present an alternative approach, based on an analytic solution of the multi-subband Boltzmann transport equation (BTE) within the effective mass approximation for the conduction electrons. The relaxation times (RTs) are calculated self-consistently and Fermi's golden rule (FGR) is used to obtain the scattering rates \cite{moors2014resistivity}. The FGR matrix elements only rely on (statistical) physical parameters of the scattering mechanisms, such as the shape, density and energy barrier strength for GBs and the standard deviation and correlation length for SR. The matrix element contribution from GBs is based on the MS model \cite{mayadas1970electrical}, while the SR contribution originates from Ando's model \cite{ando1982electronic}. We do not narrow down the SR description to the infinite barrier limit, also known as the Prange-Nee (PN) approximation, because it neglects the wave function penetration in the barrier and this approximation becomes worse when the wire dimensions decrease. Instead, we solve the matrix elements of Ando's model analytically, without expanding them for small SR sizes, but rather making use of SR distribution functions on a finite domain.

\section{Model}
\begin{figure}[!htb]
\centering
\subfigure[]{\includegraphics[width=0.3\linewidth]{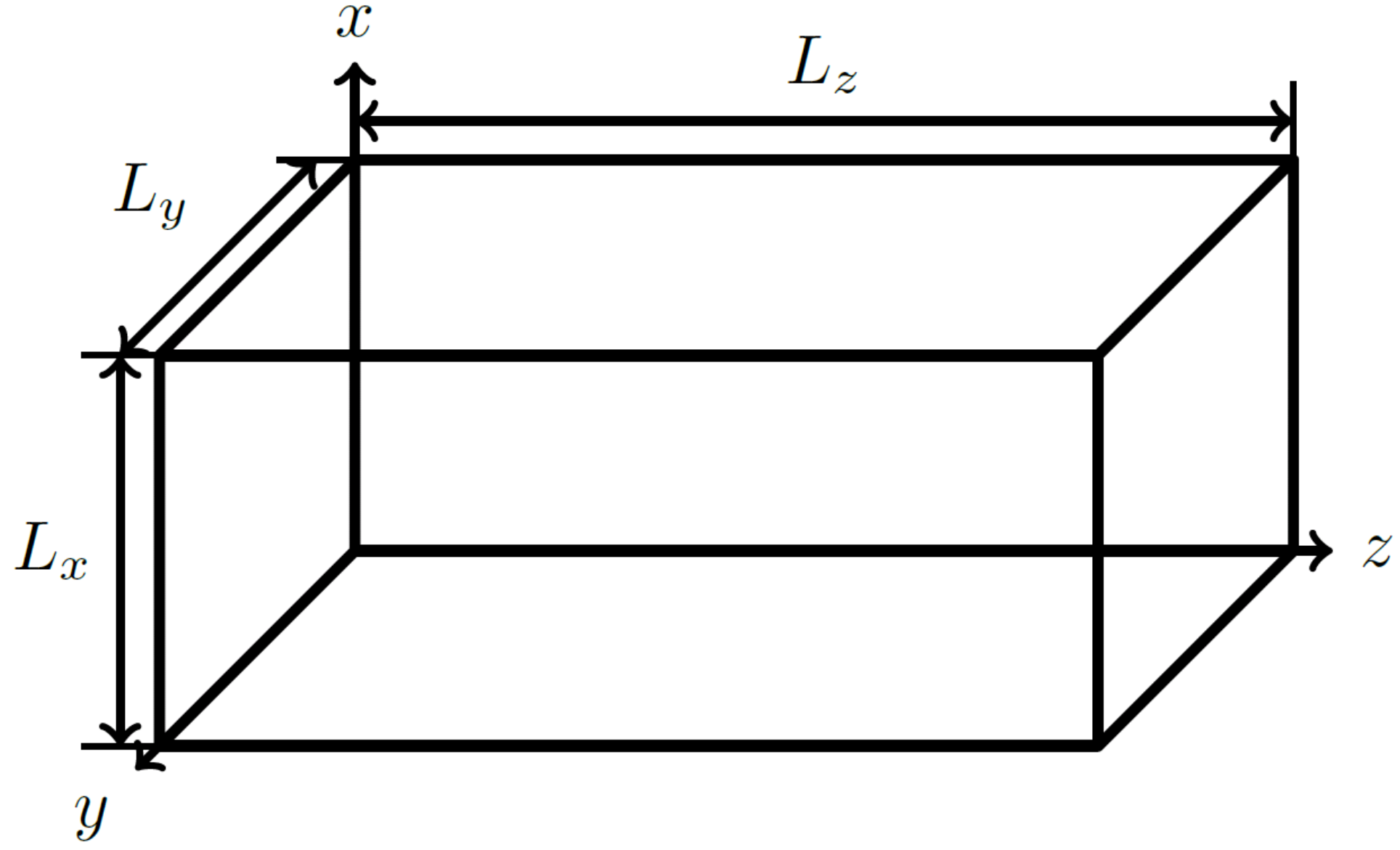}}
\subfigure[]{\includegraphics[width=0.3\linewidth]{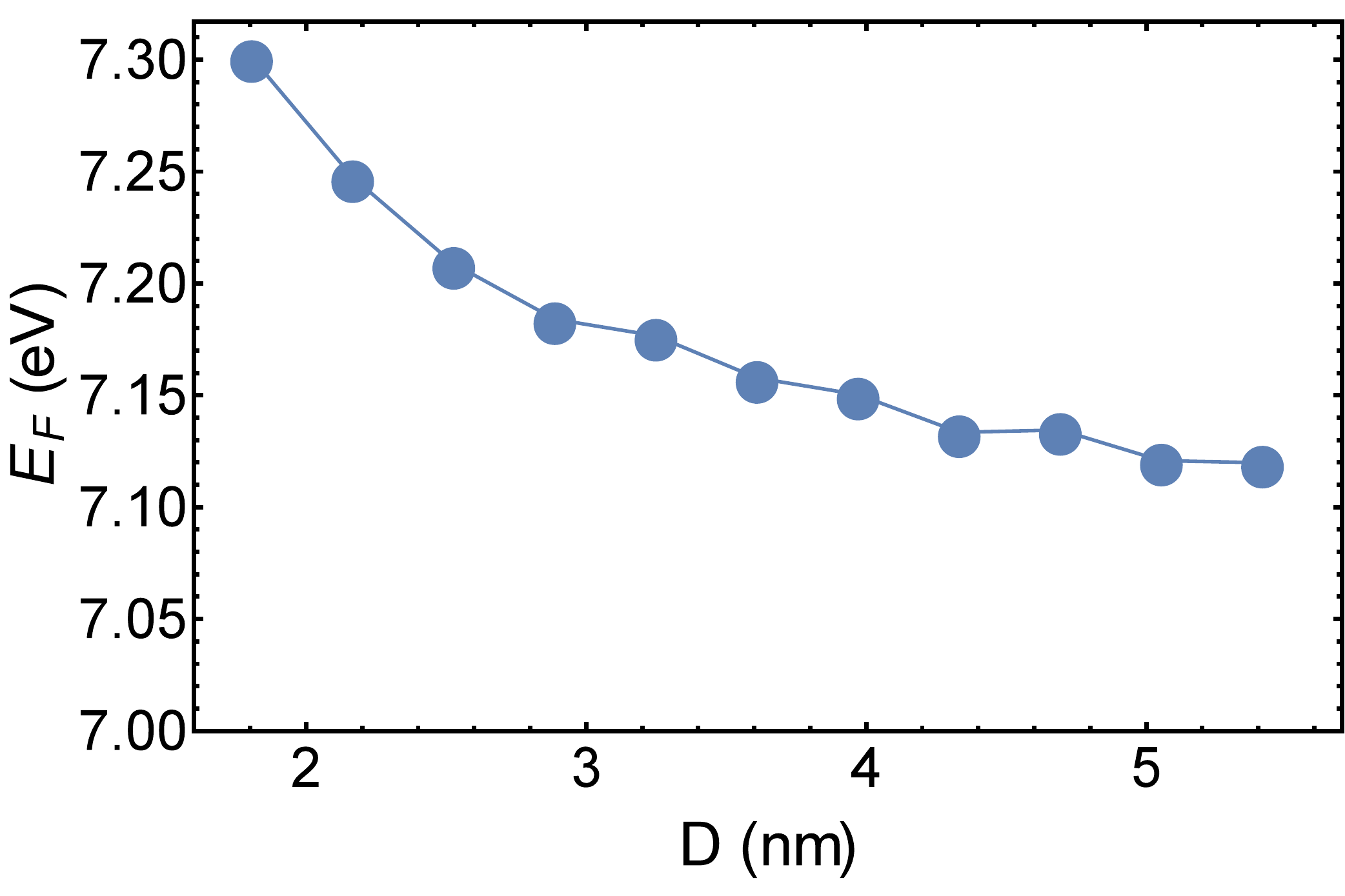}} \\
\subfigure[]{\includegraphics[width=0.3\linewidth]{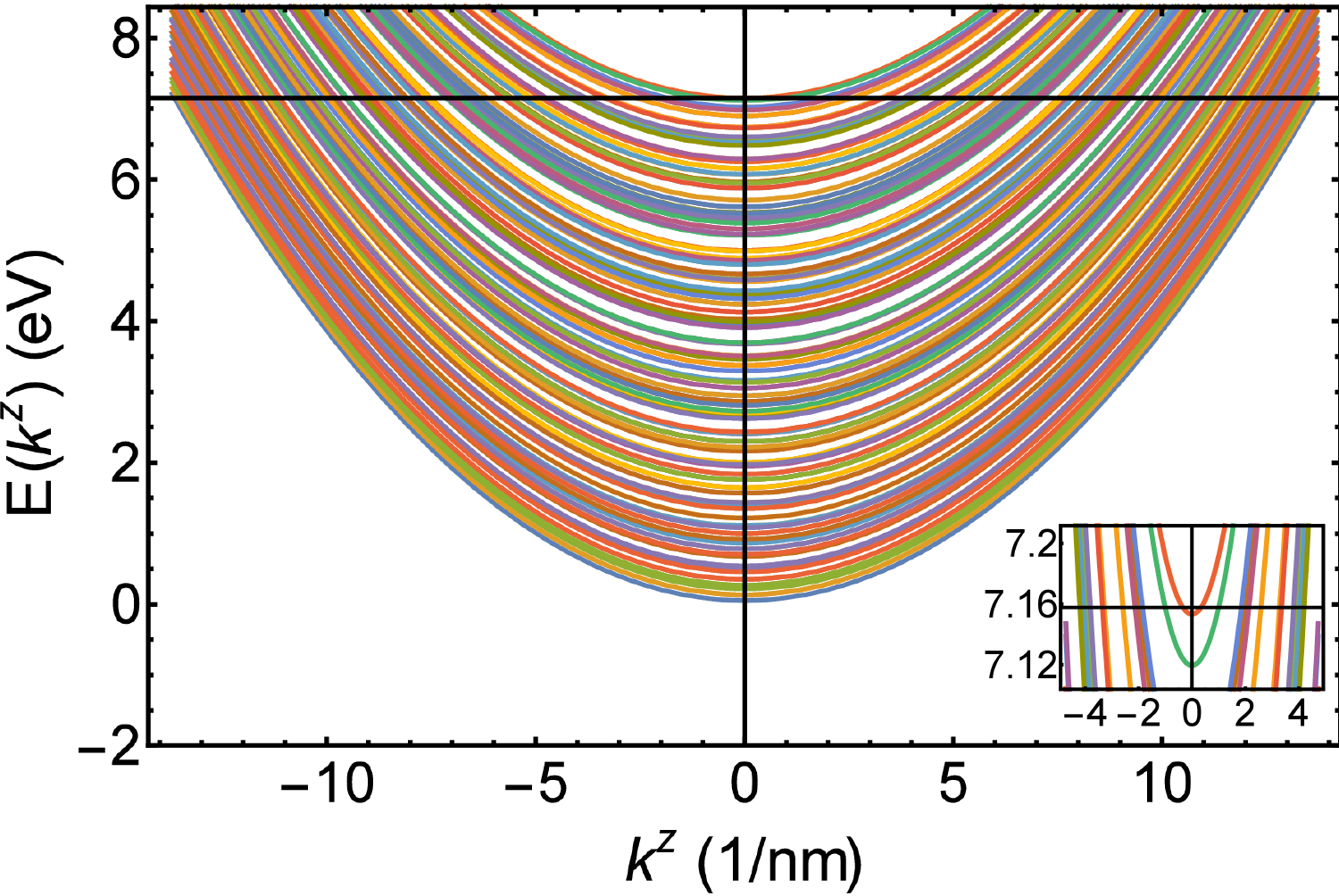}}
\subfigure[]{\includegraphics[width=0.3\linewidth]{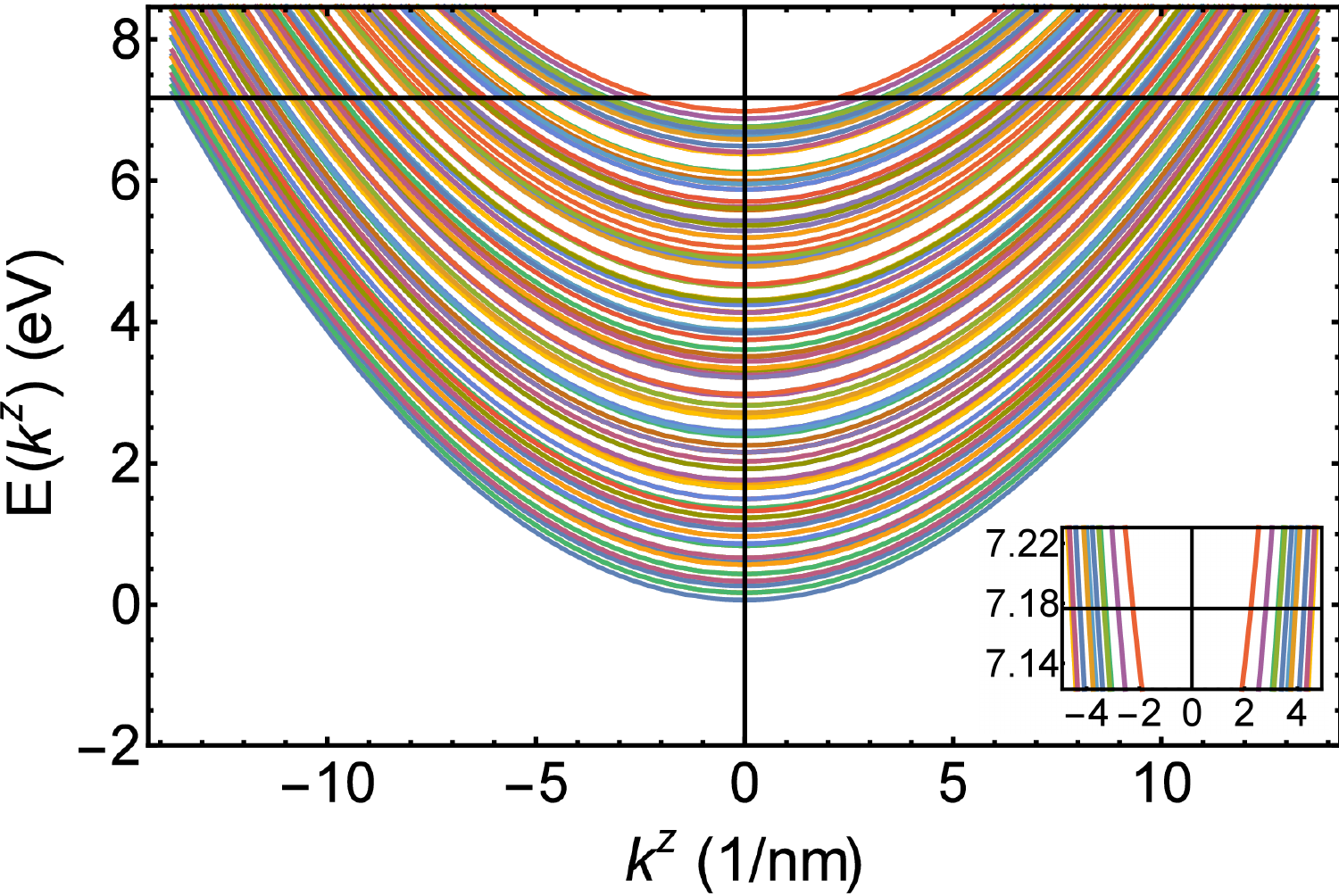}}
\caption{The unperturbed nanowire model, shown in (a), is used to calculate the Fermi level, as shown in (b), using the quantized subbands to fill up the system to reach the proper electron density of the considered metal. The subbands for two nanowires with square crosssection are shown below, corresponding to (c) $D\approx$~3.6~nm (d) $D\approx$~3.25~nm simulation result in Fig.~\ref{figSR}. Note the gap between positive and negative wave vector states at the Fermi level in (d).}
\label{figBands}
\end{figure}

The model presented here is based on the subband dependent BTE and FGR for the scattering rates. The effective mass approximation for the conduction electrons is assumed, together with zero temperature and a small electric field, such that the following system of equations has to be solved to obtain the resistivity or conductivity:
\begin{align}
\label{RTEq}
\frac{1}{\tau_i} = \frac{m_e L_z}{\hbar^3} \sum_{f} \frac{\langle \left| \langle i \mid V \mid f \rangle \right|^2 \rangle_{\tiny V}}{| k_f^z |} \lef 1 - \frac{k^z_f \tau_f }{k^z_i \tau_i} \rig,
\end{align}
with $\tau$ the RT of an electron state at the (size-dependent) Fermi level (see Fig.~\ref{figBands}), $m_e$ the (effective) electron mass, $k^z$ the wave vector along the transport direction ($z$) and $\langle \ldots \rangle_{\tiny V}$ an ensemble average over the considered scattering potentials (not to be confused with bra- and ket-state notation). The resistivity is then given by:
\begin{align}
\label{condEq}
\frac{1}{\rho} = \sigma = \frac{2e^2}{\pi m_e L_x L_y} \sum_n \tau_n \left| k^z_n \right|,
\end{align}
with $L_x/L_y$ the wire height/width and $n$ labeling the transverse subband ladders.

\subsection{Grain boundaries}
The scattering matrix element for $N$ GBs is given by:
\begin{align} \label{GBMatrixElement}
\langle i \mid V_\textnormal{\tiny GB} \mid f \rangle &= U_\textnormal{\tiny GB} L_\textnormal{\tiny GB} \mkern-5mu \int\limits_{-\infty}^{+\infty} \mkern-7mu \deriv x \mkern-5mu \int\limits_{-\infty}^{+\infty} \mkern-7mu \deriv y \; \psi^*_i (x) \psi_f (x) \psi^*_i (y) \psi_f (y) \sum_{\alpha=1}^N e^{-i (k^z_i - k^z_f) z_\alpha(x,y)},
\end{align}
with $\psi(x/y)$ the wave functions along the confinement directions, $U_\textnormal{\tiny GB} L_\textnormal{\tiny GB}$ the barrier strength of a single GB plane and $z_\alpha(x,y)$ the position of a GB plane along the transport direction. The MS model only considers GB planes that are normally oriented w.r.t. the transport direction. In this case $z_\alpha(x,y)$ is equal to a constant $z_\alpha^0$, but we also consider tilted planes that satisfy:
\begin{align}
z_\alpha(x,y) = z_\alpha^0 + \beta x + \gamma y,
\end{align}
with real parameters $\beta$ and $\gamma$. The averaging over different GB plane configurations is performed using the following GB plane distribution function:
\begin{align}
&g\lef z_1^0, \ldots, z_N^0 \rig = \frac{\exp\left[-\sum\limits_{\alpha = 1}^{N-1} \lef z_{\alpha+1}^0 - z_\alpha^0 - L_z / N \rig^2/2\sigma_\textnormal{\tiny GB}^2\right]}{L_z \lef 2\pi \sigma_\textnormal{\tiny GB}^2 \rig^{(N-1)/2}},
\end{align}
with $\sigma_\textnormal{\tiny GB}$ the GB position standard deviation. The tilt angles can be considered to be fixed or variable. In the latter case, a distribution function for the parameters $\beta, \gamma$ is also required. In this work a uniform distribution of $\beta \in \left[ - \Delta_\beta, + \Delta_\beta \right]$ with tilt parameter $\Delta_\beta$ is considered with $\gamma=0$, such that the total averaging procedure can be done analytically.

\subsection{Surface roughness}
The matrix element for SR of the $x=0$ -boundary is given by:
\begin{align} \label{SRMatrixElement}
\langle i \mid V^{x=0}_\textnormal{\tiny SR} \mid f \rangle &= \frac{U}{L_z} \mkern-5mu \int\limits_{-\infty}^{+\infty} \mkern-5mu \deriv y \mkern-10mu \int\limits_{-L_z/2}^{+L_z/2} \mkern-13mu \deriv z \; \psi^*_i (y) \psi_f (y) e^{-i (k^z_i - k^z_f) z} \int\limits_{0}^{S_{x=0}(y,z)} \mkern-23mu \deriv x \; \psi^*_i (x) \psi_f (x),
\end{align}
with $U$ the height of the potential well outside the nanowire relative to its interior and $S_{x=0}$ a SR function giving the height of the rough nanowire boundary w.r.t. the smooth unperturbed wire boundary. The PN approximation consists of an expansion of the expression for small SR sizes, such that the integration over $x$ can be approximated to be proportional to $S_{x=0}$, and an infinite potential well limit. In the expression for the scattering rate a product of two SR functions appears and the averaging procedure consists of replacing the SR function product by its expectation value:
\begin{align}
\label{gaussianCorr}
\left< S(\mathbf{r}) S(\mathbf{r}') \right> &= \Delta^2 C(\mathbf{r}, \mathbf{r}'), \qquad \qquad C(\mathbf{r}, \mathbf{r}') \equiv \exp\left[ -\frac{|\mathbf{r} - \mathbf{r}'|^2}{\Lambda^2/2} \right],
\end{align}
with $\mathbf{r},\mathbf{r}'$ two positions on the unperturbed boundary. It can be questioned whether the expansion for small SR sizes works well, because the wave functions in the confinement directions might oscillate heavily. Instead of expanding the integral with the SR function, we average over the full integral with a distribution function for the SR height at two different positions: $f(S,S')$. The natural choice, a bivariate normal distribution function, 
\begin{align}
\label{bivariateNormal}
f(S,S') & = \mkern-5mu \int\limits_{-\infty}^{+\infty} \mkern-7mu \deriv S \mkern-5mu \int\limits_{-\infty}^{+\infty} \mkern-7mu \deriv S' \; \frac{1}{2\pi \Delta^2 \sqrt{1-C\lef \mathbf{r}, \mathbf{r}' \rig^2}} \exp\left[ -\frac{S^2 + (S')^2  - 2 C\lef \mathbf{r}, \mathbf{r}' \rig S S'}{2 \Delta^2 \left[ 1 - C\lef \mathbf{r}, \mathbf{r}' \rig^2 \right]} \right],
\end{align}
does not allow for an analytical solution \cite{lizzit2014new}. Therefore we propose two distribution functions on a finite domain (FD 1 and 2) that allow for an analytical treatment of the scattering rates:
\begin{align}
\notag
f_\textnormal{FD 1}(S,S') &= \left\{ \begin{matrix}
                   1 + 4C(\mathbf{r},\mathbf{r}')/3, \quad \textnormal{if } \textnormal{sign}(S) = \textnormal{sign}(S') \; \; \; \\
                   1 - 4C(\mathbf{r},\mathbf{r}')/3, \quad \textnormal{if } \textnormal{sign}(S) = -\textnormal{sign}(S')
                  \end{matrix} \right., \\
f_\textnormal{FD 2}(S,S') &= 1 - C(\mathbf{r}, \mathbf{r}') + \delta \lef S - S' \rig C(\mathbf{r}, \mathbf{r}'),
\end{align}
while $|S|,|S'| < \sqrt{12} \Delta$ and zero otherwise, hence the name distribution functions on a finite domain. It can be easily checked that both distribution functions satisfy Eq.~\ref{gaussianCorr}.

\section{Simulation results}
All simulation results are obtained for copper nanowires with square crosssection $L_x=L_y=D$, conduction electron density $n_e =8.469\times 10^{28}$ m${}^{-3}$, bulk Fermi level $E_\textnormal{\tiny F} = 7$~eV and lattice constant $a_\textnormal{\tiny Cu}=0.361$~nm.

\subsection{Grain boundaries \& surface roughness}
\begin{figure}[!htb]
\centering
\subfigure[]{\includegraphics[width=.3\linewidth]{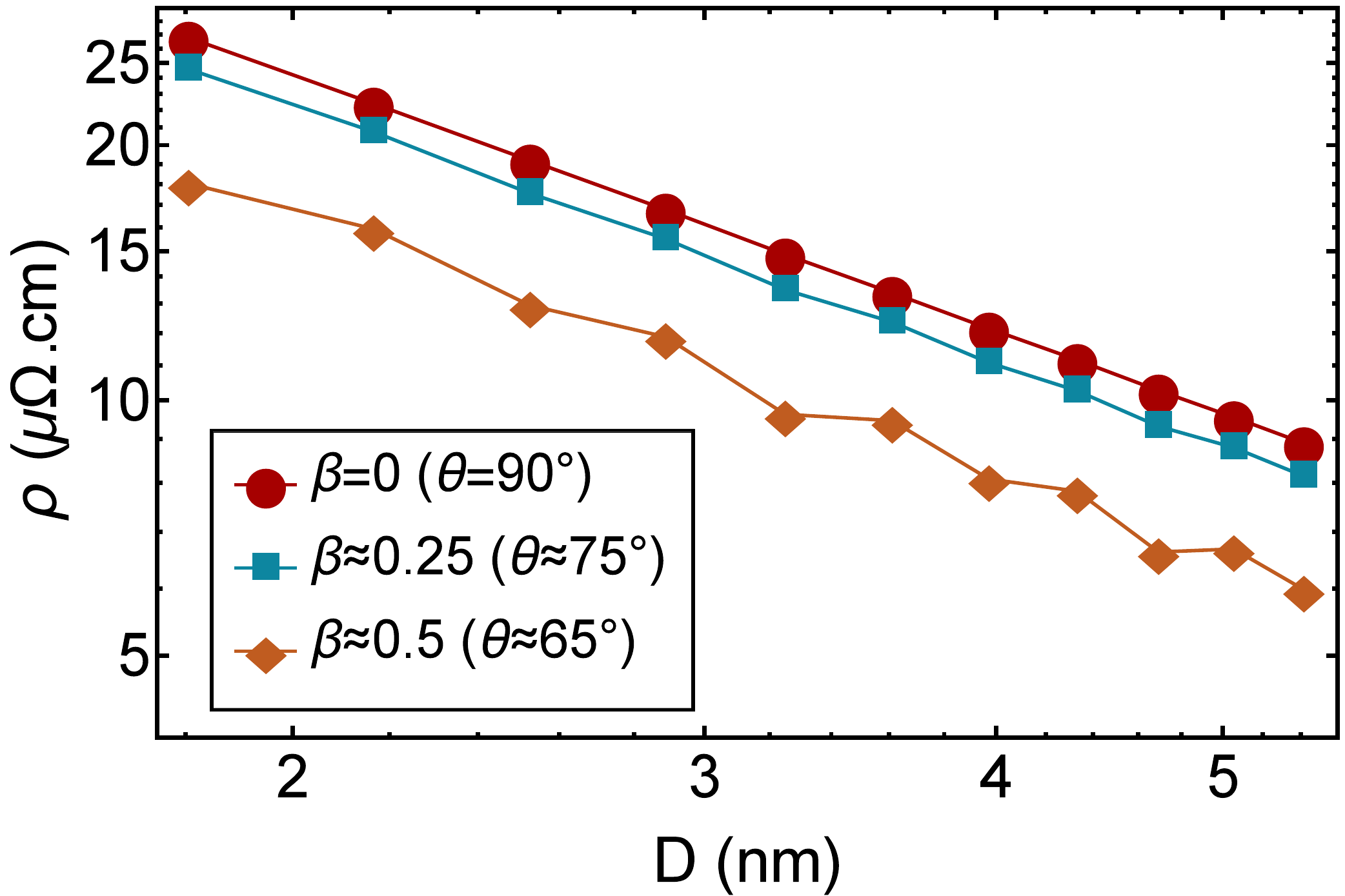}}
\subfigure[]{\includegraphics[width=.3\linewidth]{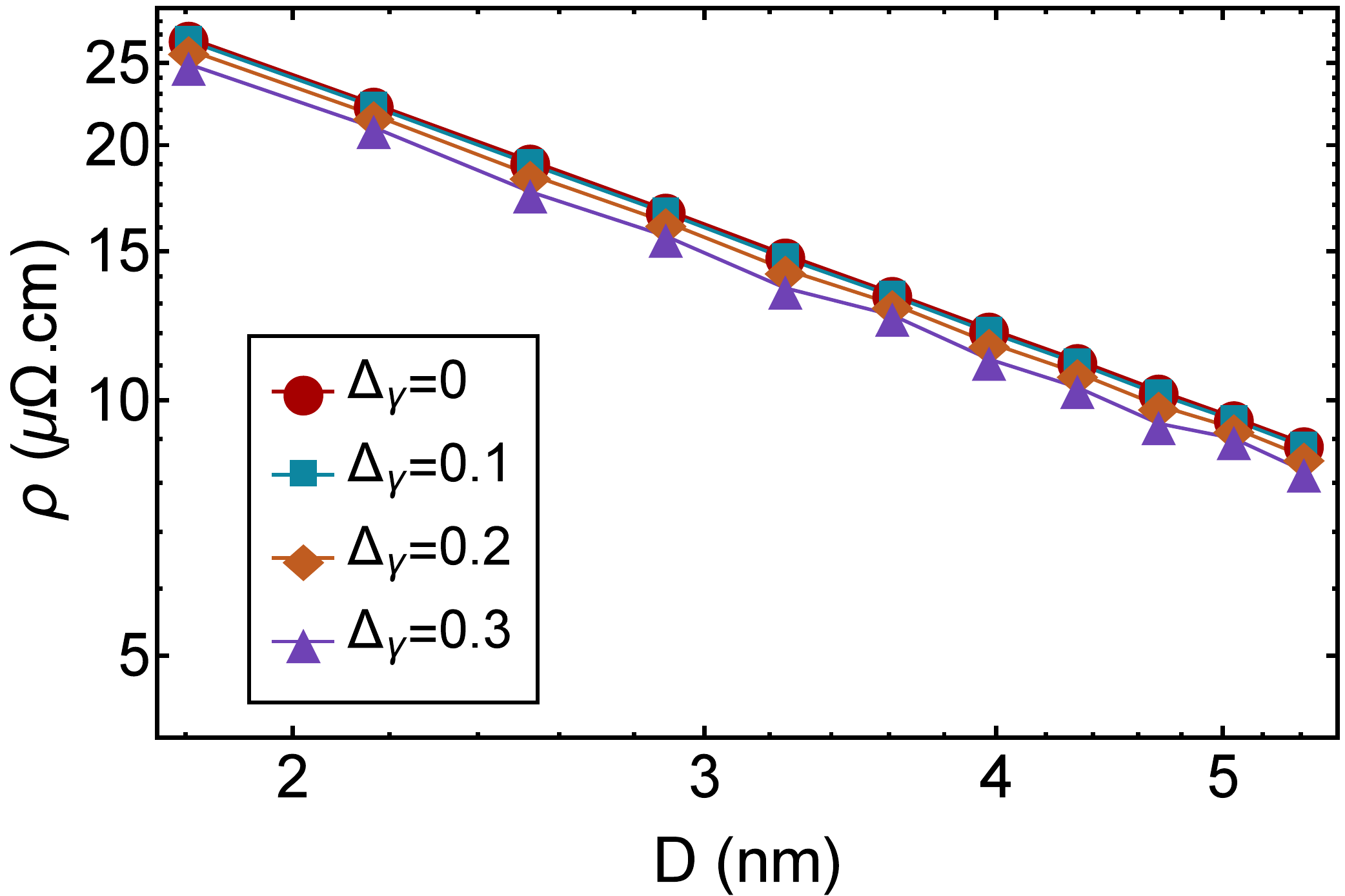}} \\
\subfigure[]{\includegraphics[width=.3\linewidth]{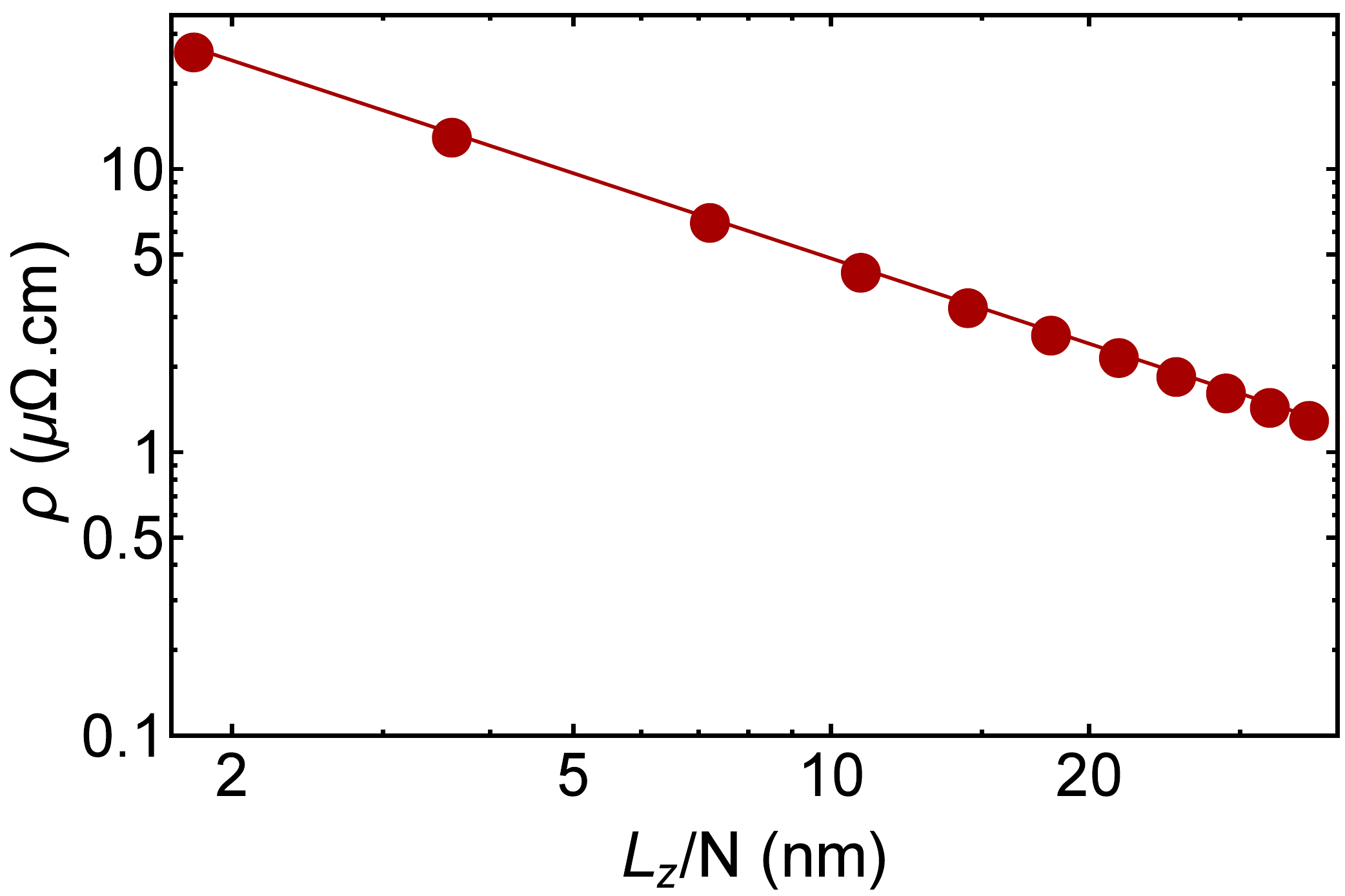}}
\subfigure[]{\includegraphics[width=.3\linewidth]{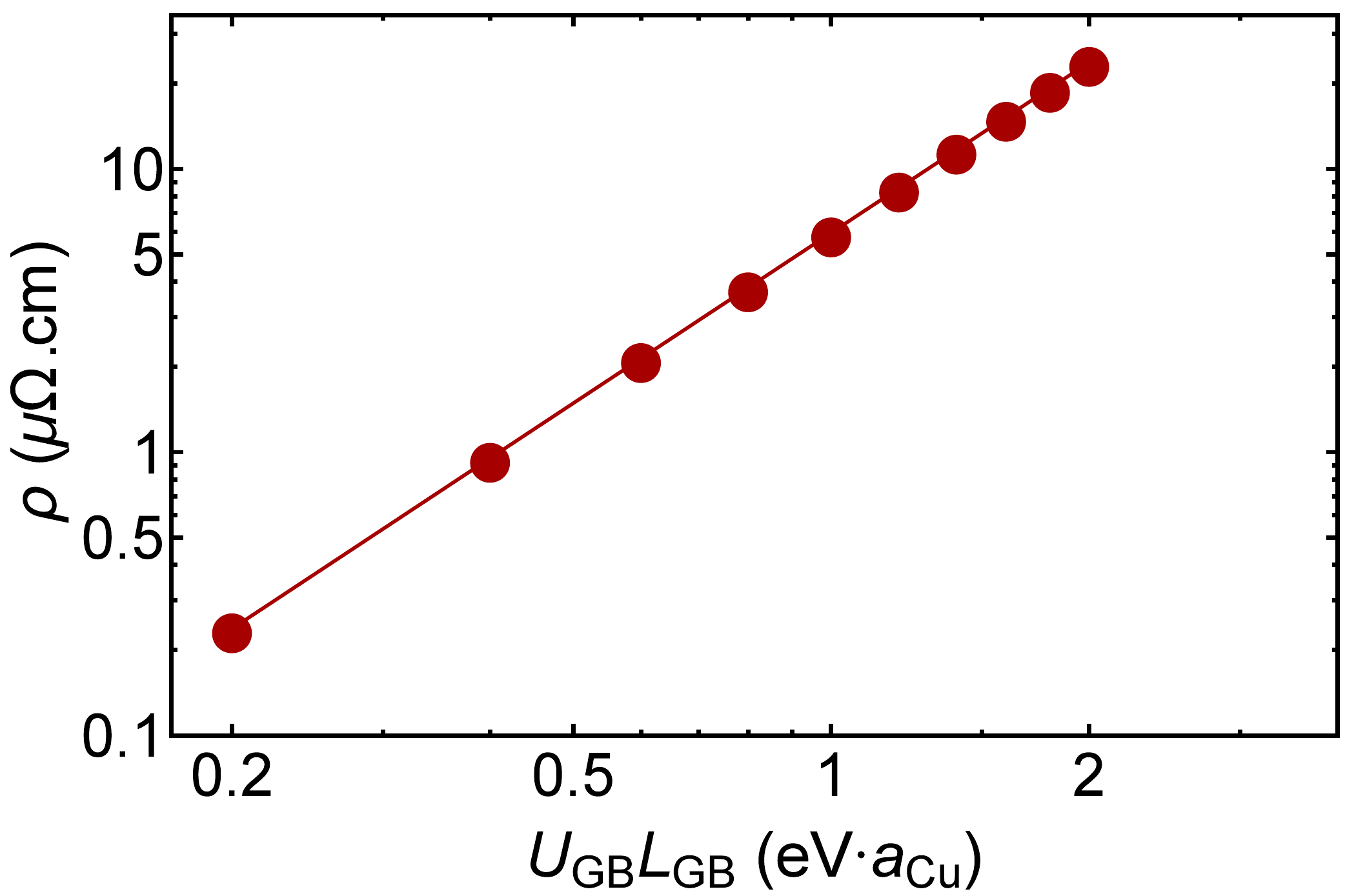}}
\caption{The resistivity contribution due to GB scattering is shown with (a) all GB planes tilted under an angle $\theta = \textnormal{arccot}(\beta)$ ($\gamma=0$) w.r.t. the $x=0$ -boundary and (b) all GB planes tilted under a random angle (tilt parameter $\Delta_\beta$) \cite{moors2015electron}). In (c) a $D\approx 3.6$~nm nanowire with non-tilted GB planes is simulated for different values of $L_z/N$ and the same nanowire is simulated for different values of $U_\textnormal{\tiny GB} L_\textnormal{\tiny GB}$ in (d). The remaining input parameters in (a-d), if not varying, are fixed at: $U_\textnormal{\tiny GB} L_\textnormal{\tiny GB} =1.5$~eV$\cdot a_\textnormal{\tiny Cu}$, $L_z/N=D$, $\sigma_\textnormal{\tiny GB} = D/4$.}
\label{figTilt}
\end{figure}

In Fig.~\ref{figTilt} the GB resistivity contribution is shown and it follows a $\rho \propto 1/L_{x/y}$ scaling. This scaling comes from the size-dependent average distance between two GB planes, $L_z/N$, considered to be proportional to $L_{x/y}$. This type of dependence on $L_z/N$ typically appears with today's processing techniques\cite{josell2009size}. When the GB planes are uniformly tilted, away from the perpendicular orientation of the MS model, the resistivity decreases substantially (see Fig.~\ref{figTilt} (a)), while the decrease is much smaller if the GB plane tilt is random (see Fig.~\ref{figTilt} (b)) \cite{moors2015electron}. Subband effects are only visible in the uniformly tilted case when the deviation from the normal MS orientation is large enough, resulting in small deviations from the $\rho \propto 1/L_{x/y}$ behavior.
The resistivity scaling due to average distance between GB planes $L_z/N$ and GB barrier strength $U_\textnormal{\tiny GB} L_\textnormal{\tiny GB}$ is shown in Fig.~\ref{figTilt}~(c-d) and obeys $\rho \propto N/L_z$, $\rho \propto (U_\textnormal{\tiny GB} L_\textnormal{\tiny GB})^2$. A change of standard deviation $\sigma_{\textnormal{\tiny GB}}$ has negligible impact on the resistivity, as long as it is of the same order of magnitude as $L_z/N$. In the limit $\sigma_{\textnormal{\tiny GB}}\rightarrow 0$, a perfect periodic superlattice is achieved and the resistivity goes to zero as expected, but this situation is unrealistic for random GBs and is therefore not considered in this work.

\begin{figure}[!htb]
\centering
\subfigure[]{\includegraphics[width=.3\linewidth]{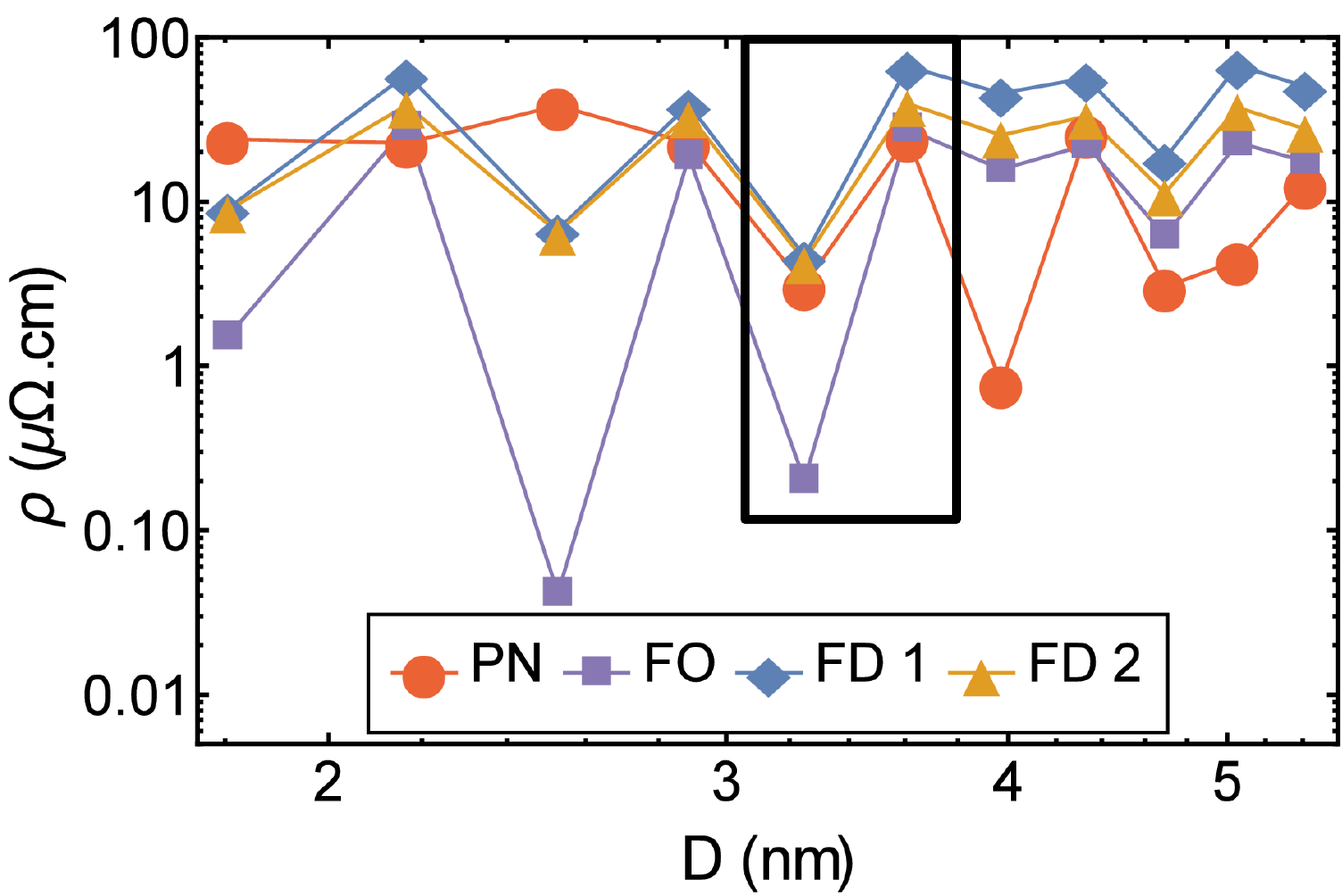}}
\subfigure[]{\includegraphics[width=.11\linewidth]{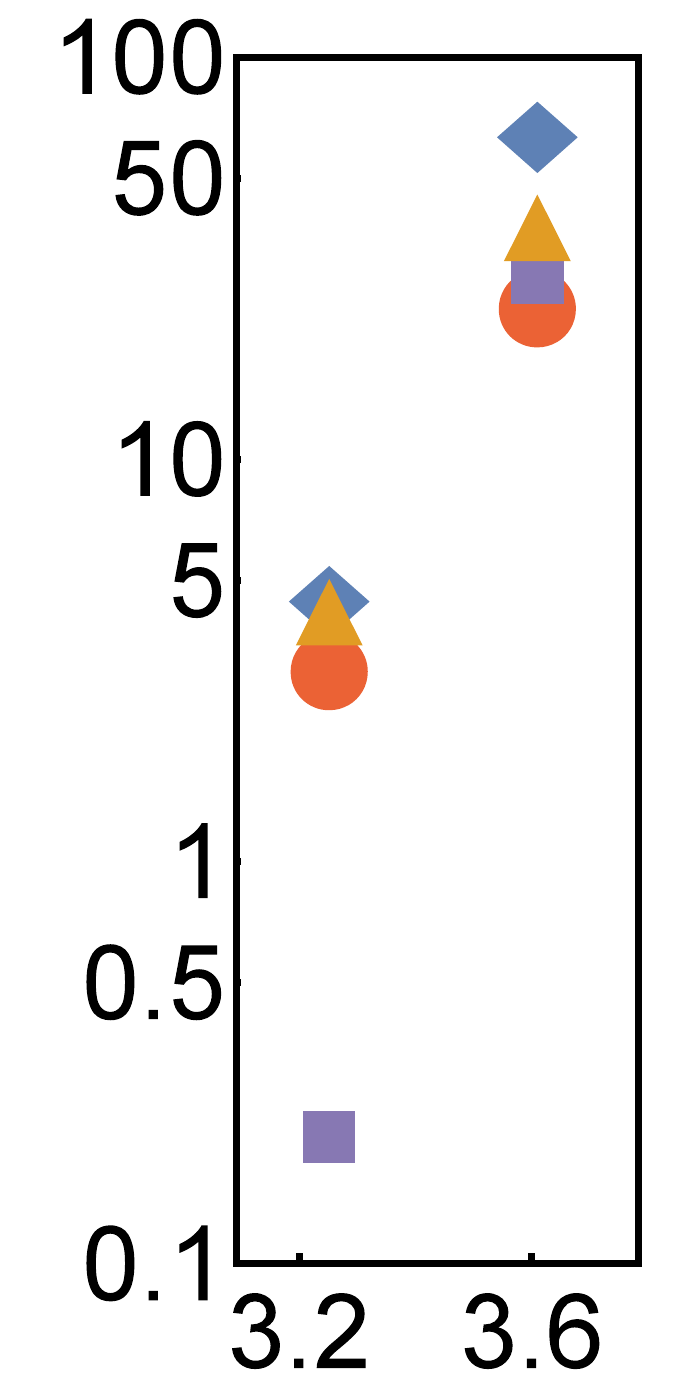}} \\
\subfigure[]{\includegraphics[width=.3\linewidth]{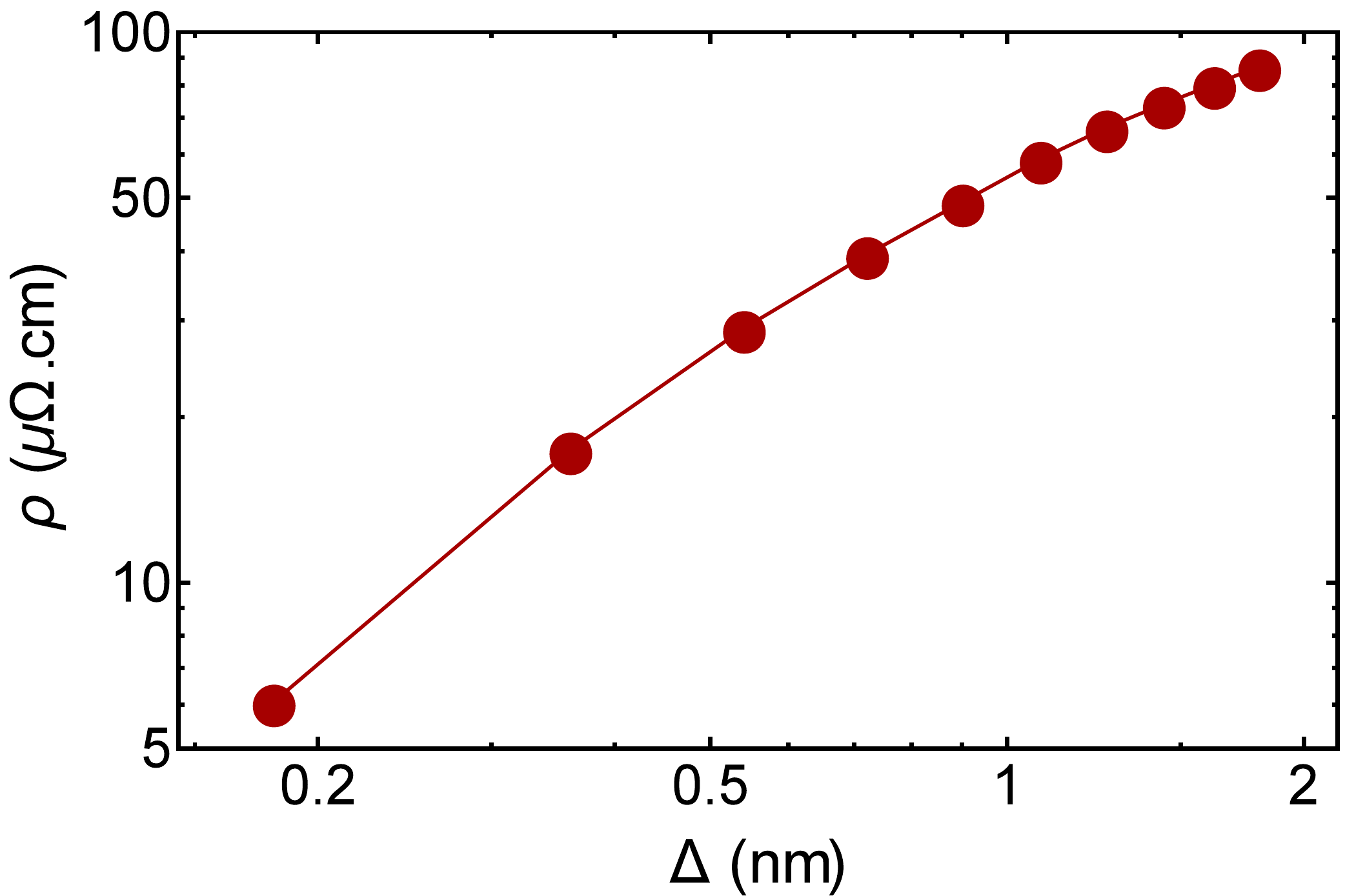}}
\subfigure[]{\includegraphics[width=.3\linewidth]{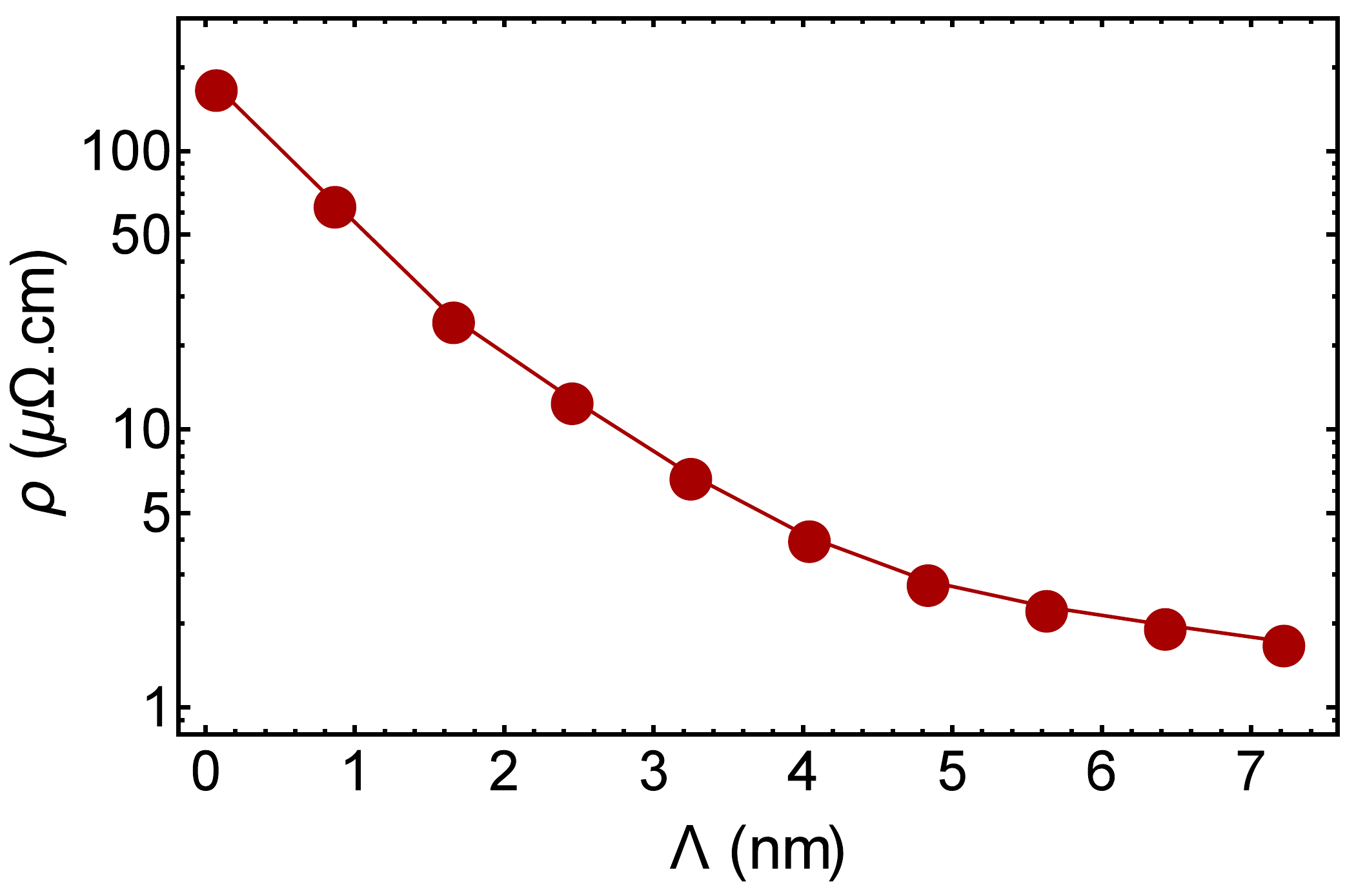}}
\caption{The resistivity contribution from SR scattering is shown in (a-b) using the PN approximation (PN), first order approximation (FO) and the two methods with finite domain distribution functions (FD 1 and 2). The wire with $D\approx 3.25$~nm has a larger than average momentum gap (Fig.~\ref{figBands}~(b)) and a much lower resistivity than the other nanowires. The dependence on $\Delta$ and $\Lambda$ is shown in (c) and (d) for a $D\approx 3.6$~nm nanowire with use of FD 2.}
\label{figSR}
\end{figure}

For SR scattering, the newly introduced finite domain distribution functions are compared with a first order and PN approximation of Eq.~\ref{SRMatrixElement} in Fig.~\ref{figSR}~(a), showing good agreement of the two new methods while PN and the first order approximation show substantial deviations. The subband quantization has a large impact on the SR resistivity contribution which makes it difficult to extract a simple resistivity scaling law as a function of the height or width. The dependence on SR standard deviation $\Delta$ and correlation length $\Lambda$ is slightly clearer and shown in Fig.~\ref{figSR} (c-d), although simple power law scaling is not observed. Even if the standard deviation is an important parameter and has an important influence on the resistivity, the correlation length has an even larger impact. The resistivity is exponentially suppressed when the correlation length increases.

\subsection{Self-consistent relaxation times \& Matthiessen rule}
\begin{figure}[!htb]
\centering
\subfigure[]{\includegraphics[width=.3\linewidth]{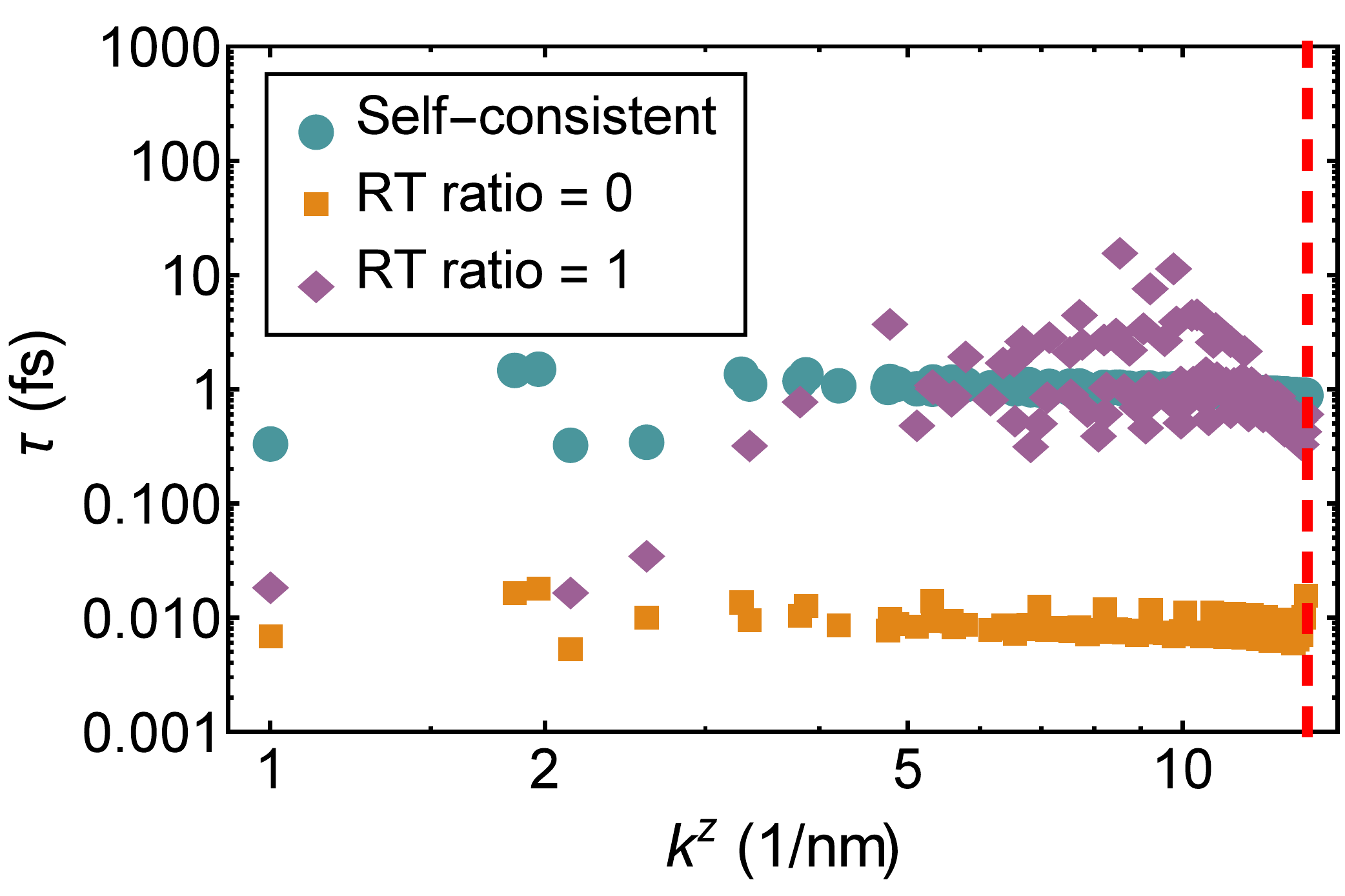}}
\subfigure[]{\includegraphics[width=.3\linewidth]{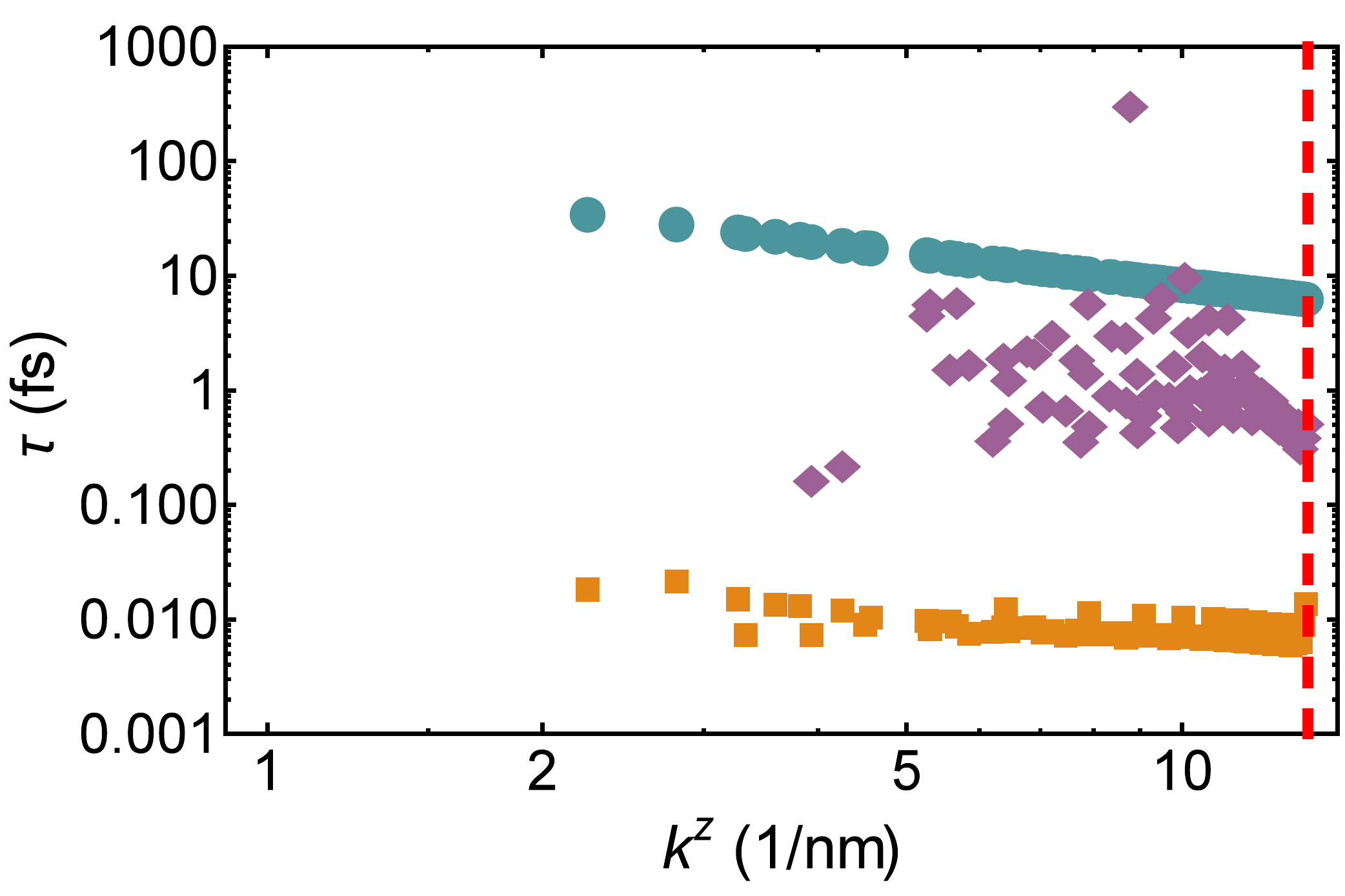}} \\
\subfigure[]{\includegraphics[width=.3\linewidth]{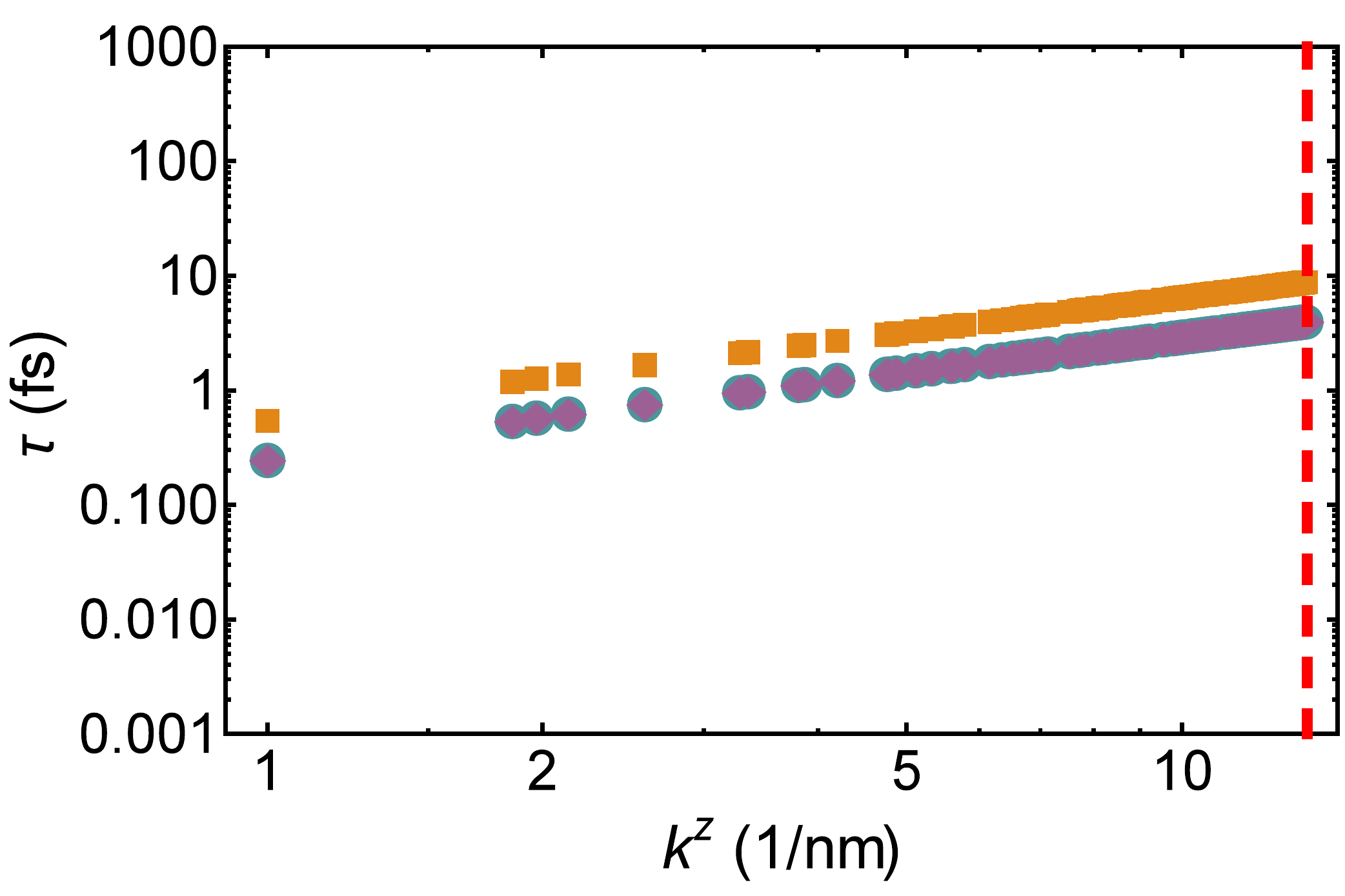}}
\subfigure[]{\includegraphics[width=.3\linewidth]{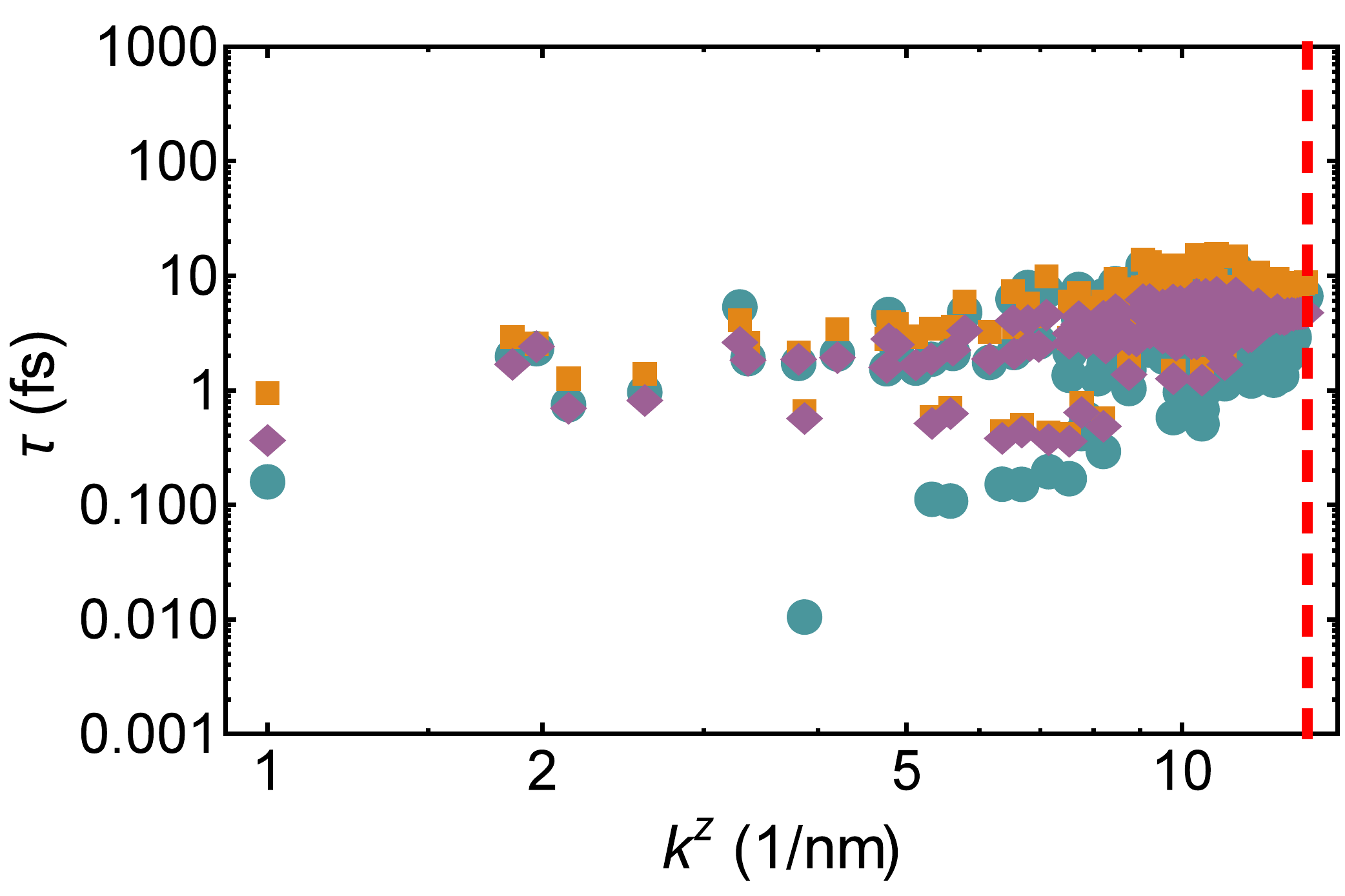}}
\caption{The RTs of the electron states at the Fermi level are given as a function of the wave vector (along the transport direction) for (a) SR scattering with small momentum gap, $D\approx 3.6$~nm (b) SR scattering with large momentum gap, $D\approx 3.25$~nm (c) GB scattering with normally oriented GB planes (d) GB scattering with uniformly tilted GB planes, $\beta\approx 0.5$, $\theta\approx 64^\circ$. The RTs of the multi-subband BTE are shown for a self-consistent solution and the solutions with RT ratio on the right-hand side of Eq.~\ref{RTEq} replaced by zero (RT ratio = 0) and one (RT ratio = 1).}
\label{figRT}
\end{figure}

\begin{figure}[!htb]
\centering
\subfigure[]{\includegraphics[width=.35\linewidth]{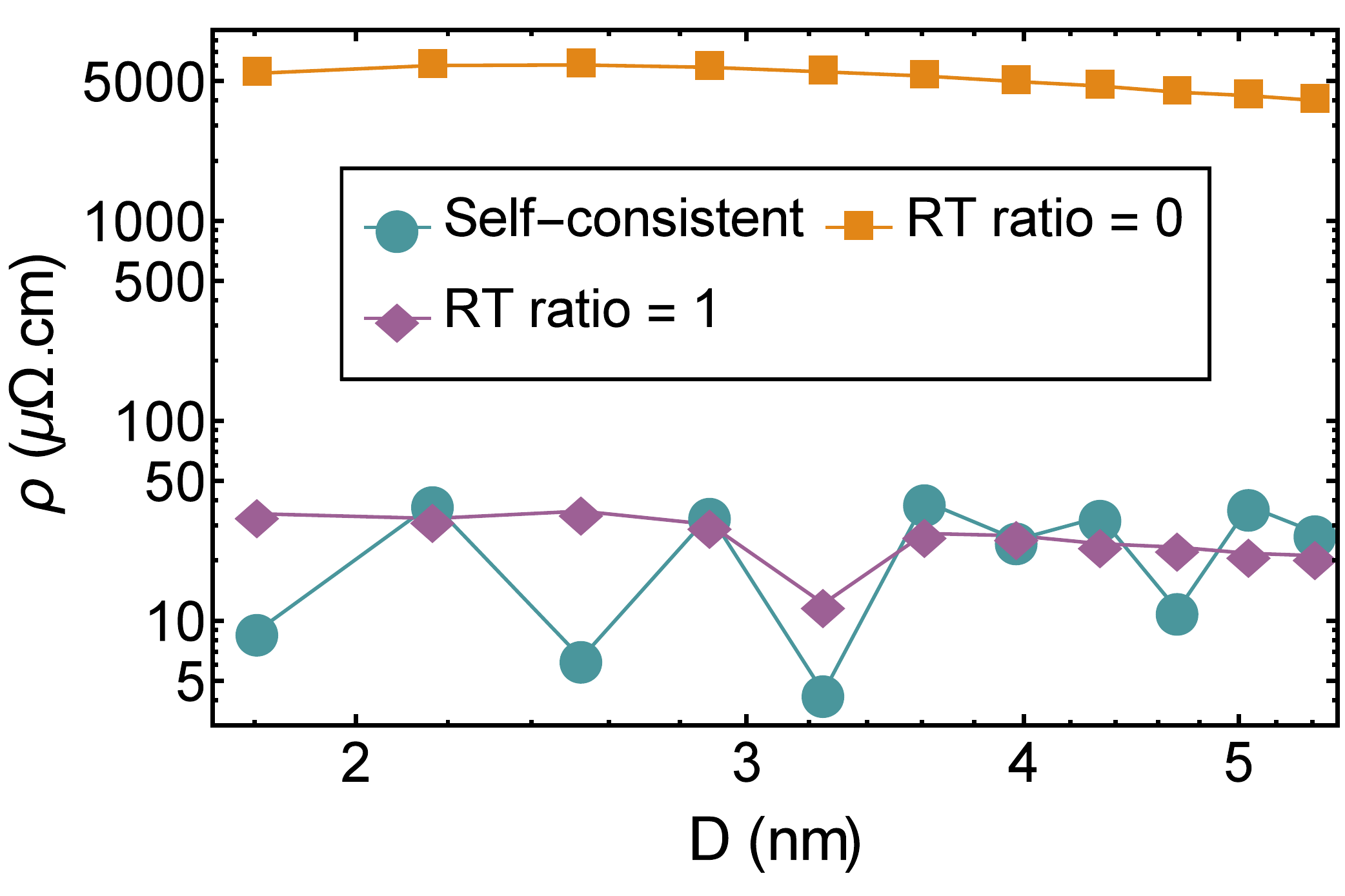}}
\subfigure[]{\includegraphics[width=.35\linewidth]{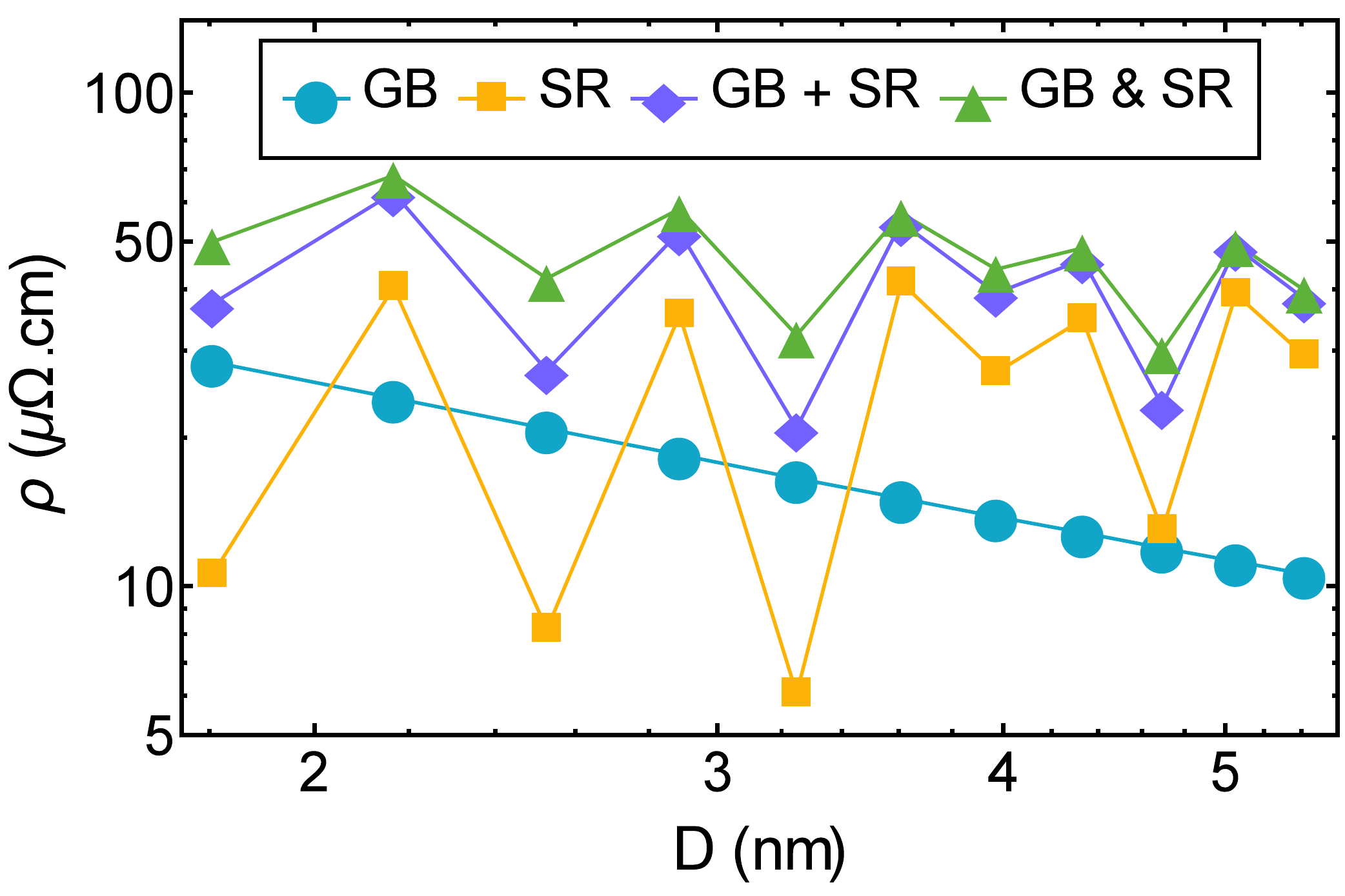}}
\caption{In (a) the resistivity scaling for a nanowire due to SR scattering is shown for a self-consistent solution and the solutions with RT ratio on the right-hand side of Eq.~\ref{RTEq} replaced by zero (RT ratio = 0) and one (RT ratio = 1). In (b) different scalings are shown, when only GB scattering is turned on (GB), only SR scattering turned on (SR), both of them turned on (SR\&GB) and adding the resistivities together (SR+GB). The bulk resistivity has also been added in all cases. The parameters in both figures are: $U_\textnormal{\tiny GB} L_\textnormal{\tiny GB} =1.5$~eV$\cdot a_\textnormal{\tiny Cu}$, $L_z/N=D$, $\sigma_\textnormal{\tiny GB} = D/4$ (in (a) only SR); $\Delta = 2a_\textnormal{\tiny Cu}\approx 0.7$~nm, $\Lambda = 5a_\textnormal{\tiny Cu}\approx 1.8$~nm, use of distribution function FD 2.}
\label{figMRule}
\end{figure}

An important remark concerning the multi-subband BTE method is that the RTs should be calculated self-consistently to obtain correct results. Approximating the RT ratio of initial and final state on the right-hand side of Eq.~\ref{RTEq} by one or even zero often fails to give a good result in a nanowire (see Fig.~\ref{figRT} and Fig.~\ref{figMRule}). Note that replacing the ratio by one works fine for the MS model (without GB plane tilt) because there is only scattering to the electron state from the same subband but with the transport wave vector being flipped, for which the RT is exactly the same due to reflection symmetry in the system. Dropping the RT ratio alltogether makes a huge underestimation of the RTs in the case of SR, while they are overestimated in the case of GB scattering.

The Matthiessen rule that is often invoked to obtain the total resistivity can also be tested with the model presented here. In Fig.~\ref{figMRule} one can see the difference between adding the resistivities separately due to GB and SR scattering at the end and solving for the resistivity by including both scattering mechanisms in the collision term in the beginning and solving for the RTs self-consistently.

\subsection{Highly conductive nanowires}
The different methods for SR scattering show resistivity values that are up to an order of magnitude smaller than the typical values. The reason for this appears to be a momentum/wave vector gap of the electron states at the Fermi level. For SR scattering, transitions to states with lowest difference in wave vector are dominant, thereby suppressing back-scattering when such a gap (of the order of 1/nm) is present. The subbands of the $D\approx 3.6$~nm and $D\approx 3.25$~nm data points are shown in Fig.~\ref{figBands} (c-d) and their SR resisivity contribution in Fig.~\ref{figSR} (a-b). The nanowire with wave vector gap has a much lower resistivity, confirmed by all methods. Interestingly, the low resistivity value is only visible in the self-consistent RT solution, as can be seen in Fig.~\ref{figMRule}~(a). Finally, the effect remains visible when the GB scattering and bulk resistivity contributions are included (see Fig.~\ref{figMRule}~(b)).

\section{Conclusion}
In this work a model, based on the multi-subband BTE is presented that is able to retrieve the resistivity scaling in metal nanowires, including the quantum-mechanical scattering and subband quantization effects and without making use of phenomenological parameters. Both GB and SR scattering are taken into account and the MS model for GBs is generalized to tilted GB planes while the description of SR, based on Ando's model, is more accurate than PN or other approximate methods with the use of finite domain distribution functions. The scattering rates for GBs and SR both have analytical expressions which allows for simulations of metal wires with realistic dimensions and having a large number of subbands.

The simulations show a general increasing trend in resistivity for smaller wire dimensions, but the inverse scaling with width, predicted by FS and MS models, is not valid in general. For GB scattering, the normal GB plane orientation causes Umklapp scattering, the worst case scenario due to full current reversal, resulting in a maximal resistivity contribution from GB scattering. The resistivity is lowered substantially when the GB planes are all tilted under the same angle and slightly lowered when each GB plane is randomly tilted. Crucial parameters are the GB barrier strength and average distance between two GB planes. The behavior of SR scattering is more subjected to quantization effects and resistivity scaling as a function of height or width is difficult to extract. Both the SR standard deviation and correlation length have an important impact on the resistivity, which increases for larger standard deviation and smaller correlation length.

To tackle the detrimental resistivity scaling in metal nanowires, GB scattering could be reduced significantly by reducing the GB barrier strength, increasing the average distance between two GB planes or inducing an overall tilt of the GB planes. The resistivity due to SR is hard to tackle, as small correlation lengths have the largest contribution. Even if the quality of the nanowire increases and the standard deviation can be kept under control, the exponential dependence on the correlation length would still cause a substantial contribution from intrinsic atomic-scale SR. However, by tuning the subbands at the Fermi level in such a way that left- and right-movers have a substantial gap between them, the SR contribution can be largely suppressed.

\bibliographystyle{IEEEtran}

\bibliography{IEEEabrv,bare_confBibTeX}

\end{document}